\documentclass[11pt]{article}

\usepackage[utf8]{inputenc}
\usepackage[T1]{fontenc}
\usepackage[letterpaper,margin=1in]{geometry}
\usepackage{amsmath}
\usepackage{newtxtext,newtxmath}
\usepackage{microtype}
\usepackage{graphicx}
\usepackage{booktabs}
\usepackage{siunitx}
\usepackage{xcolor}
\usepackage[font=small,labelfont=bf,labelsep=period,skip=6pt]{caption}
\usepackage{titlesec}
\usepackage{fancyhdr}
\usepackage{enumitem}
\usepackage[numbers,square,sort&compress]{natbib}
\usepackage[colorlinks=true,linkcolor=linkblue,citecolor=linkblue,urlcolor=linkblue]{hyperref}
\usepackage[capitalise,nameinlink]{cleveref}

\definecolor{linkblue}{RGB}{0,72,160}
\setlist{nosep}

\titleformat{\section}{\normalfont\large\bfseries}{\thesection}{0.75em}{}
\titleformat{\subsection}{\normalfont\normalsize\bfseries}{\thesubsection}{0.75em}{}
\titlespacing*{\section}{0pt}{1.4em}{0.5em}
\titlespacing*{\subsection}{0pt}{1.0em}{0.3em}

\fancypagestyle{firstpage}{%
  \fancyhf{}
  \fancyfoot[C]{\thepage}
}

\newcommand{\stdric}{Standard RIC}
\newcommand{\ancoff}{ANC RIC\ensuremath{_{\text{ANC off}}}}
\newcommand{\ancon}{ANC RIC\ensuremath{_{\text{ANC on}}}}

\hypersetup{
  pdftitle={Active noise cancellation extends the benefit of AI speech enhancement in real hearing-aid recordings},
  pdfauthor={Andrew T. Sabin, Linkai Li},
}

\begin{document}
\thispagestyle{firstpage}

% ---------------------------------------------------------------------------
% Title block
% ---------------------------------------------------------------------------
\begin{center}
  \rule{\linewidth}{2pt}\par\vspace{0.6em}
  {\LARGE\bfseries Active noise cancellation extends the benefit of\\[0.15em]
   AI speech enhancement in real hearing-aid recordings\par}
  \vspace{0.5em}\rule{\linewidth}{0.6pt}\par
  \vspace{1.6em}
  \begin{tabular}{c@{\hspace{4em}}c}
    \textbf{Andrew T. Sabin}\textsuperscript{*} & \textbf{Linkai Li} \\
    Bose Corporation & Orka \\
    \href{mailto:Andy_Sabin@bose.com}{\texttt{Andy\_Sabin@bose.com}} & \href{mailto:linkai@hiorka.com}{\texttt{linkai@hiorka.com}} \\
  \end{tabular}\par
  \vspace{1.2em}
  {\small\scshape A preprint -- \today}\par
  \vspace{0.4em}
  {\footnotesize\textsuperscript{*}Corresponding author}
\end{center}
\vspace{0.8em}

\begin{center}
  \textbf{\large Abstract}
\end{center}
\begin{quote}
\small
Active noise cancellation (ANC) is common in headphones but rare in hearing aids, and its value for understanding speech in noise is not established. Algorithms that improve signal-to-noise ratio (SNR) act only on the hearing aid's amplified path, while unprocessed sound also reaches the eardrum directly through the vent, limiting the SNR improvement that is realized. Our earlier simulations~\citep{sabin2024} predicted that attenuating this direct path with ANC would substantially increase the benefit of such processing. We tested that prediction in three hardware conditions: a standard receiver-in-canal (RIC) hearing aid, and an ANC-capable RIC with ANC off and on. Each was fit to NAL-NL2 targets for an N3 hearing loss and recorded on KEMAR with an AI speech-enhancement algorithm (AI-SE) off and on, in seven realistic scenes above \SI{70}{dB} SPL. Speech and noise spectra were estimated from each recording with a binary time--frequency mask derived from the clean speech. AI-SE produced a similar peak benefit of about \SI{4}{dB} in the high frequencies in all three conditions, but that benefit stayed at or above \SI{1}{dB} only above 1500, 600, and \SI{130}{Hz} for the \stdric, \ancoff, and \ancon, respectively, closely tracking the measured attenuation of the direct path. With AI-SE on, \ancon{} improved SII-weighted SNR by \SI{5.1}{dB} relative to the open ear, compared with \SI{3.3}{dB} for \ancoff{} and \SI{1.6}{dB} for the \stdric. With AI-SE off, enabling ANC added only \SI{0.2}{dB}. ANC is therefore an enabling technology: on its own it added little, but by attenuating the direct path it allowed SNR-enhancing processing to deliver substantially greater benefit at the eardrum, extending that benefit to much lower frequencies.
\end{quote}

\vspace{0.3em}

% ---------------------------------------------------------------------------
\section{Introduction}
% ---------------------------------------------------------------------------
Hearing aids improve speech understanding in noise by applying a wide range of algorithms to the signal delivered by the receiver. Any SNR improvement these algorithms create exists in that signal. The sound at the eardrum, however, is the sum of the amplified path and a direct path: sound that enters the ear canal through the vent or around the ear coupling without passing through the hearing aid~\citep{sabin2024,winkler2016}. As processing reduces the noise in the amplified path, the unprocessed direct-path noise becomes an increasingly large fraction of the total noise at the eardrum and sets a ceiling on the SNR that can be realized at the eardrum~\citep{sabin2024,serizel2010}. This ceiling is most consequential at lower frequencies, where venting is most transparent and prescribed gain is typically smallest.

Active noise cancellation (ANC) offers a way to raise that ceiling. Throughout this paper, ANC refers specifically to attenuation of direct-path sound by destructive interference, not to the gain-based ``noise reduction'' algorithms common in hearing aid amplified paths. ANC has been used in hearing devices to reduce the occlusion effect~\citep{mejia2008} and has been analyzed in integrated ANC and noise-reduction systems~\citep{serizel2010,serizel2013}. Two properties of ANC matter here. First, ANC attenuates speech and noise arriving through the direct path alike, so it does not by itself change the direct-path SNR. Second, by attenuating the direct path it allows an SNR improvement created in the amplified path to dominate at the eardrum over a wider frequency range. Any intelligibility benefit of ANC in noise is therefore expected to be indirect and to depend on the level of gain and the performance level of SNR-enhancing processing.

Our previous simulation study formalized this mechanism~\citep{sabin2024}. Modeled ANC benefit increased with environmental sound level, with the SNR improvement the device could generate, with ANC strength, and with better hearing sensitivity. For a moderate hearing loss, the model predicted substantial gains in intelligibility metrics in environments above \SI{70}{dB} SPL when the device improved SNR by more than about \SI{4}{dB} SNR. Those simulations isolated ANC from every other device property but did not involve physical hardware.

The present study evaluates this mechanism in recordings of commercial receiver-in-canal (RIC) hearing aids. We compared a standard RIC (Orka O1) with the Orka O1 Pro, an ANC-capable RIC, recorded with ANC disabled and enabled. Each hardware condition was recorded with a proprietary AI speech-enhancement algorithm (AI-SE) off and on. We asked (1)~how much SNR each condition added at the eardrum relative to the open ear, (2)~whether ANC extended the AI-SE benefit to lower frequencies, and (3)~how these acoustic changes translated into SII-weighted SNR.

% ---------------------------------------------------------------------------
\section{Methods}
% ---------------------------------------------------------------------------
\subsection{Design and device conditions}
The design consisted of three hardware conditions each with two processing states. The hardware conditions were a standard RIC hearing aid (``\stdric''; Orka O1) and an ANC-capable RIC hearing aid (Orka O1 Pro) with ANC disabled (``\ancoff'') or enabled (``\ancon''). Within each hardware condition, AI-SE was off or on, giving six total aided conditions. Both devices ran the same hearing-aid signal-processing suite and were programmed to the same targets (see below), so differences among hardware conditions are attributable to the receiver, its acoustic coupling, or the ANC state.

The two devices differ in the acoustics of the receiver and its coupling to the ear. The \stdric{} uses a balanced-armature receiver with a tulip dome. The ANC-capable RIC uses a custom moving-coil receiver with a closed dome and a vent in the plastic housing of the ear-canal component. Comparisons between \stdric{} and \ancoff{} therefore reflect combined differences in transducer and coupling. In contrast, \ancoff{} and \ancon{} were recorded with the same device, fitting, prescribed gain, and signal-processing settings, so ANC state was the only intended difference between them. Differences in direct paths were quantified via real-ear occluded insertion gain (REOIG), measured with the device in place and its amplified path muted.

All aided conditions were programmed for the Bisgaard N3 standard audiogram~\citep{bisgaard2010} using experienced user NAL-NL2 targets~\citep{keidser2011}, so that differences among conditions were not attributable to different prescribed gains. Real-ear insertion gain (REIG) was estimated from the recordings as the across-recording mean ratio of aided to open-ear mixture spectra, smoothed over one-half octave.

\subsection{Acoustic scenes and recordings}
Recordings were made on a KEMAR manikin placed at the center of a ring of eight loudspeakers in a quiet room, following the HearAdvisor laboratory protocol~\citep{sabin2023}. Background scenes were drawn from the Ambisonic Recordings of Typical Environments (ARTE) database~\citep{weisser2019}. This report uses the seven ARTE scenes with overall noise levels above \SI{70}{dB} SPL (\cref{tab:scenes}), the environments in which the modeled ANC benefit is expected to be largest~\citep{sabin2024}. In each scene, six talker conditions (2 versions of scenes with 1, 2, or 3 talkers) were presented at the scene's nominal SNR, giving 42 recordings (7 scenes $\times$ 6 talker conditions) per condition. Recordings were two-channel (left and right ear simulators) at \SI{44.1}{kHz}. Each aided recording had a corresponding open-ear recording and clean speech signal.

\begin{table}[t]
  \centering
  \caption{Acoustic scenes used in the analysis.}
  \label{tab:scenes}
  \begin{tabular}{l S[table-format=2.0] S[table-format=-1.1]}
    \toprule
    Scene & {Level (dB SPL)} & {Nominal SNR (dB)} \\
    \midrule
    Cafe 1         & 71 & 2.9 \\
    Cafe 2         & 72 & 3.4 \\
    Dinner Party   & 73 & 2.2 \\
    Street Balcony & 74 & 1.1 \\
    Train Station  & 77 & 0.1 \\
    Food Court 1   & 78 & -0.4 \\
    Food Court 2   & 80 & -1.1 \\
    \midrule
    Mean           & 75 & 1.2 \\
    \bottomrule
  \end{tabular}

  \vspace{0.4em}
  \begin{minipage}{0.62\linewidth}
    \footnotesize\textit{Note.} Levels are the ARTE scene levels. Nominal SNR is the presentation SNR of the talker relative to the scene measured in the ring center with a reference mic; it is not the SNR estimated at the manikin's ears.
  \end{minipage}
\end{table}

\subsection{Estimation of speech and noise spectra}
\begin{figure}[t]
  \centering
  \includegraphics[width=\linewidth]{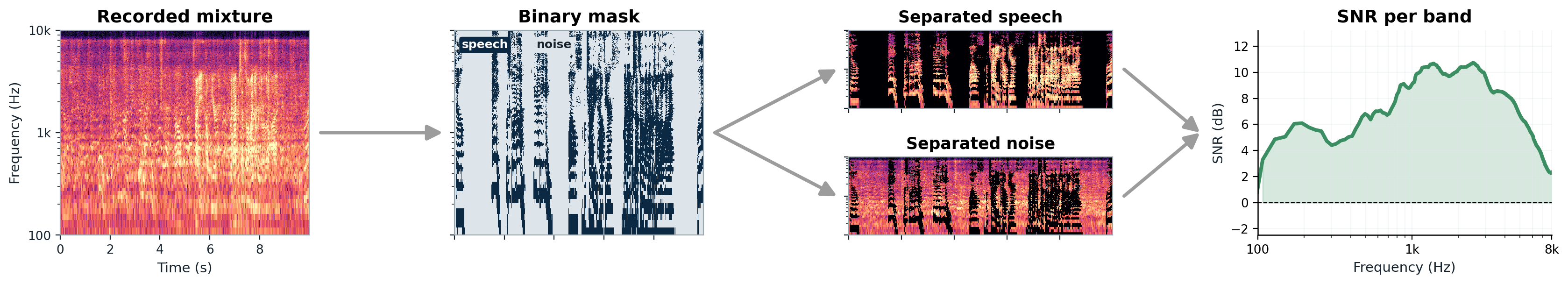}
  \caption{Estimation of speech and noise spectra from a recording. A binary mask derived from the latency-aligned clean speech reference labels each time--frequency bin of the recorded mixture as speech or noise. Power spectra of the two components yield SNR as a function of frequency. The example is one \ancon, AI-SE-on recording from the Food Court~1 scene.}
  \label{fig:method}
\end{figure}

Signal-to-noise ratio at the manikin ear was estimated via the method summarized in \cref{fig:method}. Each ear channel was analyzed separately with a short-time Fourier transform (2048-sample Hann window, 512-sample hop). Each recording was first aligned to its clean speech reference.

A binary time--frequency mask was derived from the clean speech reference. For each frequency bin, the 40th percentile of the clean reference's log power over time served as a frequency-specific floor, and a time--frequency bin was labeled ``speech'' when the clean reference exceeded that floor by at least \SI{12}{dB}; all other bins were labeled ``noise.'' The mask was then applied to the aligned device recording. For each frequency bin, the noise power was the mean recorded power over noise-labeled frames, and the speech power was the mean recorded power over speech-labeled frames minus that noise power. Speech and noise spectra were averaged across the two ears. Note that this decomposition is simply an analysis tool applied to the recordings; it is unrelated to the algorithm running inside the hearing aids.

\subsection{Outcome measures}
SNR per frequency band was computed by taking the ratio of the estimated speech and noise power spectra. For display, curves were smoothed over one octave. \emph{Added SNR} was the aided SNR minus the SNR of the matched open-ear recording; positive values indicate that the device improved relative to the open ear. \emph{AI-SE benefit} was the AI-SE-on SNR minus the AI-SE-off SNR within a hardware condition. The lower-frequency edge of the AI-SE benefit was defined as the lowest frequency above which the across-recording mean AI-SE benefit stayed at or above \SI{1}{dB}.

For a broadband summary, SNR band levels were weighted by the ANSI S3.5-1997 one-third-octave band-importance function for average speech~\citep{ansi1997}, interpolated to the Fourier bins and normalized to unit sum. \emph{Added SII-weighted SNR} is the broadband value for the aided recording minus that for the matched open-ear recording. This quantity summarizes any increase to SNR weighted by the frequency regions that matter most for speech. All summary outcomes were computed per recording and are reported as the mean and standard error of the mean (SEM) over the 42 recordings.

% ---------------------------------------------------------------------------
\section{Results}
% ---------------------------------------------------------------------------
\subsection{Gain and attenuation of the direct path}
\begin{figure}[t]
  \centering
  \includegraphics[width=\linewidth]{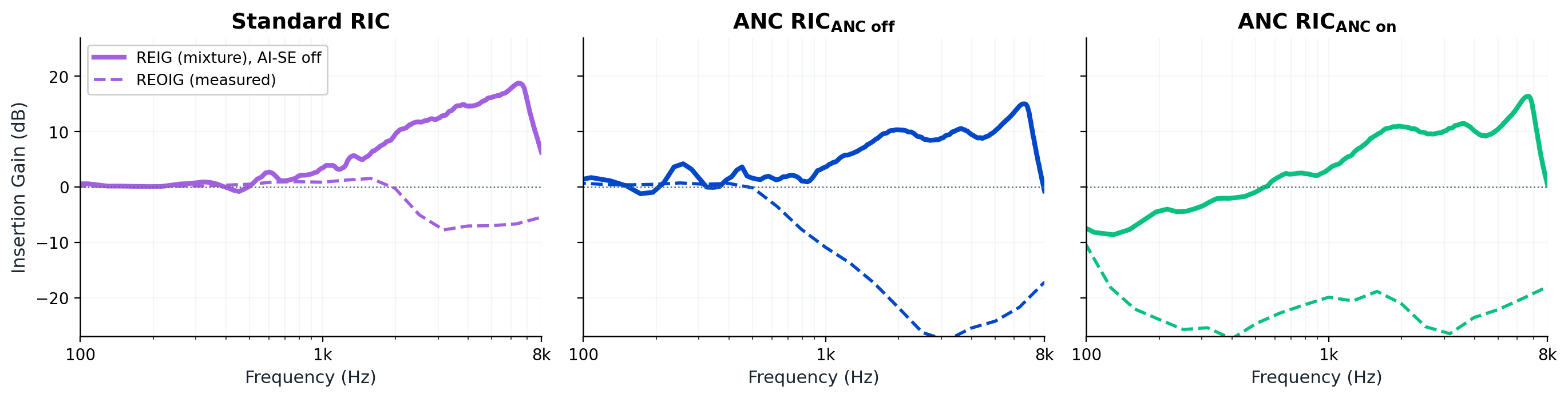}
  \caption{Amplified and direct paths by hardware condition. Solid curves: across-recording mean real-ear insertion gain (REIG) with AI-SE off averaged across all recordings and ears. Dashed curves: measured real-ear occluded insertion gain (REOIG) with the amplified path muted, an estimate of direct-path transmission. Each hardware condition is shown in one color.}
  \label{fig:reig}
\end{figure}

Average REIG was computed across all recordings and ears and is shown in \cref{fig:reig} (solid lines). For all three hardware conditions, REIG rose from near \SI{0}{dB} at low frequencies to peaks of about 15--\SI{19}{dB} near \SI{7}{kHz}, consistent with the target NAL-NL2 prescription for the N3 audiogram. While the gain profiles largely matched, the conditions differed a bit at low frequencies: \ancon{} showed negative REIG below roughly \SI{550}{Hz}, where little gain is prescribed and ANC removed part of the direct-path sound that would otherwise reach the eardrum.

The direct path (measured via REOIG) differed markedly across conditions (\cref{fig:reig}, dashed curves). For the \stdric, REOIG remained near \SI{0}{dB} up to about \SI{2}{kHz}. This REOIG indicates that the tulip dome was quite open as it transmitted low and mid frequencies essentially unattenuated. The closed coupling of the ANC-capable RIC attenuated the direct path even with ANC off, with REOIG falling below \SI{0}{dB} above about \SI{500}{Hz} and reaching about \SI{-27}{dB} near \SI{3.5}{kHz}. Enabling ANC added attenuation at the lowest frequencies: REOIG was \SI{-10}{dB} at \SI{100}{Hz} and $-20$ to \SI{-28}{dB} from \SI{200}{Hz} to \SI{1}{kHz}. The three hardware conditions therefore provided progressively stronger and broader isolation of the eardrum from the direct path.

\subsection{Added SNR across frequency}
\begin{figure}[t]
  \centering
  \includegraphics[width=\linewidth]{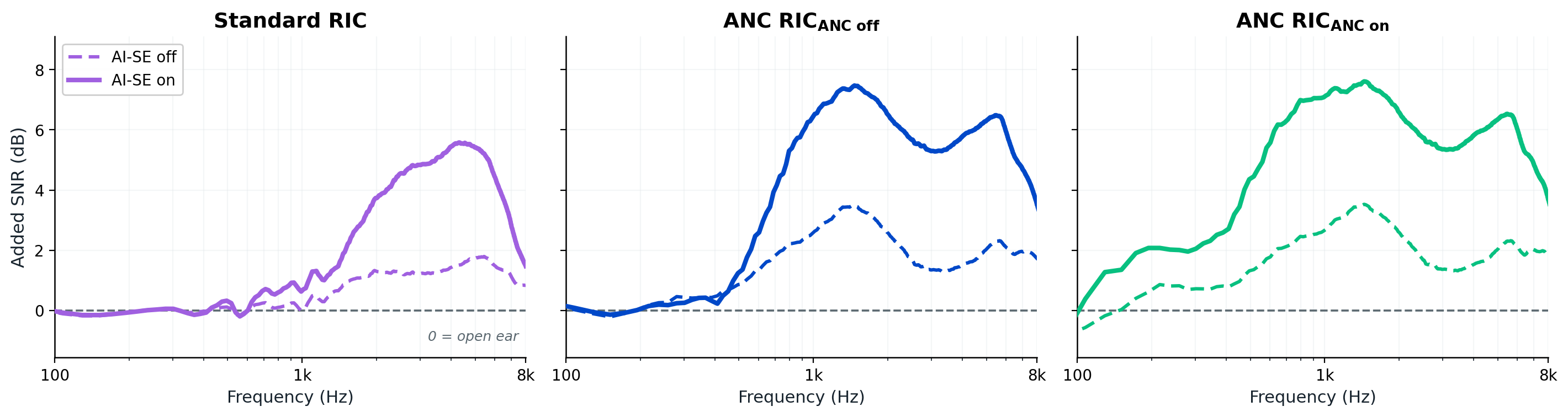}
  \caption{Added SNR relative to the open ear. Across-recording means of the aided-minus-open-ear SNR spectrum in the seven loud scenes, one panel per hardware condition. Dashed: AI-SE off; solid: AI-SE on. Curves are octave smoothed.}
  \label{fig:added}
\end{figure}

Added SNR (aided minus open-ear SNR per band) depended on both hardware and AI-SE state (\cref{fig:added}). Even with AI-SE off, all three conditions added SNR above \SI{1}{kHz} relative to the open ear (peaks of about \SI{1.8}{dB} for the \stdric{} and \SI{3.5}{dB} for both ANC RIC conditions), presumably reflecting traditional device processing (e.g., beamforming and traditional noise suppression). Turning AI-SE on increased added SNR in every condition, but over different frequency ranges. For the \stdric, the increase appeared mainly above about \SI{1}{kHz} and peaked near \SI{5.5}{dB} at 4--\SI{5}{kHz}. For \ancoff, the increase extended down to about \SI{500}{Hz} and peaked near \SI{7.5}{dB} at \SI{1.5}{kHz}. For \ancon, positive added SNR extended across nearly the full plotted range, with a similar \SI{7.5}{dB} peak.

\subsection{ANC extended the frequency range of the AI-SE benefit}
\begin{figure}[t]
  \centering
  \includegraphics[width=0.72\linewidth]{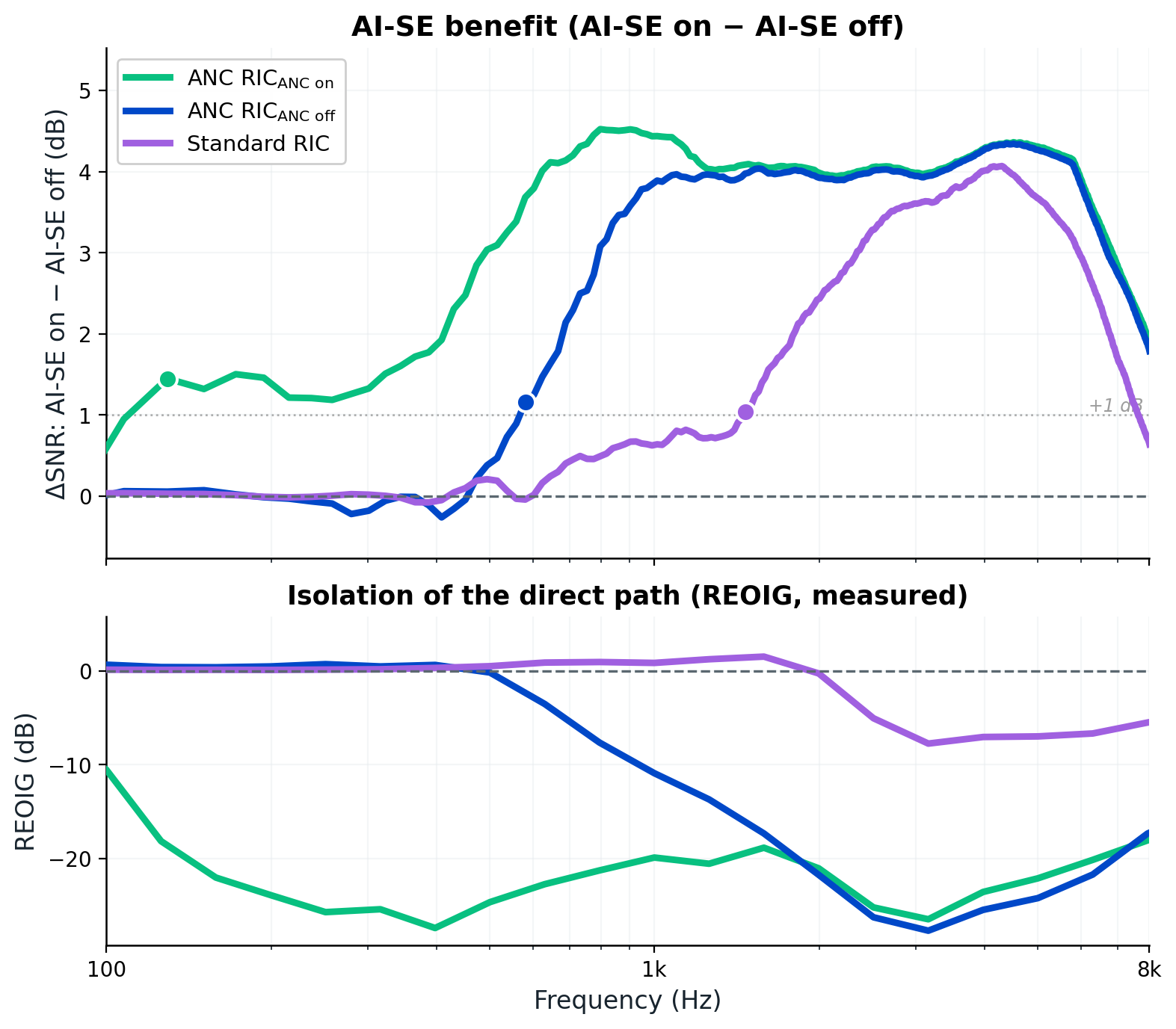}
  \caption{AI-SE benefit and direct-path attenuation. Top: across-recording mean AI-SE-on minus AI-SE-off SNR spectrum for each hardware condition. Dots mark the lowest frequency above which each curve stayed at or above \SI{1}{dB} (dotted line). Bottom: measured REOIG, replotted from \cref{fig:reig} on the same frequency axis. The lower-frequency reach of the AI-SE benefit extended as the direct path was more strongly attenuated.}
  \label{fig:benefit}
\end{figure}

Subtracting the AI-SE-off from the AI-SE-on spectrum within each hardware condition isolates the SNR change attributable to AI-SE (\cref{fig:benefit}, top). Each curve reached a similar peak of about \SI{4}{dB} between 4 and \SI{6}{kHz}, indicating that the algorithm itself produced a comparable improvement in the amplified path in each device. What differed was how far down in frequency that improvement was realized at the eardrum. The lower-frequency edge, defined as the lowest frequency above which the mean AI-SE benefit stayed at or above \SI{1}{dB}, was \SI{1500}{Hz} for \stdric, \SI{600}{Hz} for \ancoff, and \SI{130}{Hz} for \ancon. Weighted by the ANSI band-importance function, the bands at and above these edges carry approximately 53\%, 84\%, and 100\% of the importance of the speech spectrum.

The lower-frequency edges aligned well with the direct-path measurements (\cref{fig:benefit}, bottom). Where REOIG was near \SI{0}{dB}, unprocessed direct sound reached the eardrum and prevented the AI-SE improvement from appearing in the total signal. The closed coupling of the ANC-capable RIC lowered that boundary, and ANC lowered it further.

\subsection{Broadband summary}
\begin{figure}[t]
  \centering
  \includegraphics[width=0.62\linewidth]{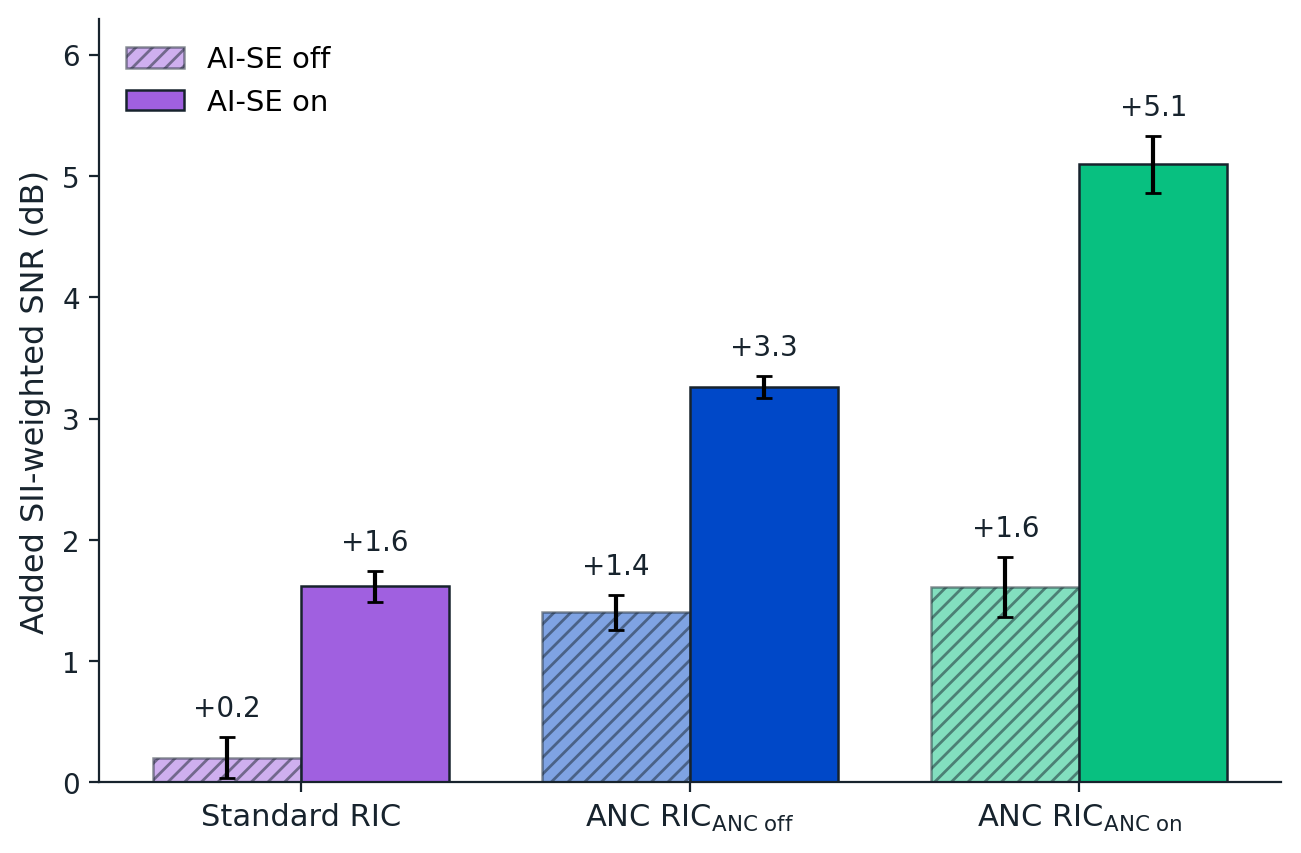}
  \caption{Broadband summary outcomes. Mean added SII-weighted SNR relative to the matched open-ear recording for each hardware $\times$ AI-SE condition. Hatched: AI-SE off; solid: AI-SE on. Error bars: SEM over 42 recordings.}
  \label{fig:summary}
\end{figure}

\Cref{fig:summary} summarizes the broadband outcome. Added SII-weighted SNR increased monotonically from \stdric{} to \ancoff{} to \ancon, and AI-SE added more in each successive hardware condition (1.4, 1.9, and \SI{3.5}{dB}). With AI-SE on, the \stdric{} added \SI{1.6}{dB} of SII-weighted SNR, whereas \ancon{} added \SI{5.1}{dB}, more than three times as much. About half of that \SI{3.5}{dB} difference (\SI{1.7}{dB}) came from the closed coupling alone (\ancoff{} with AI-SE on: \SI{3.3}{dB}), and the remainder (\SI{1.8}{dB}) came from enabling ANC. With AI-SE off, by contrast, enabling ANC changed added SII-weighted SNR by only \SI{0.2}{dB} (1.4 vs.\ \SI{1.6}{dB}).

% ---------------------------------------------------------------------------
\section{Discussion}
% ---------------------------------------------------------------------------
The central finding is that ANC was valuable only in combination with SNR-enhancing processing. With AI-SE off, enabling ANC changed added SII-weighted SNR by just \SI{0.2}{dB}. With AI-SE on, the same change added \SI{1.8}{dB}, and the full ANC-plus-AI-SE condition delivered \SI{5.1}{dB} relative to the open ear, more than three times the \SI{1.6}{dB} delivered by the \stdric{} running the same algorithm. ANC is therefore not an SNR-improving technology in its own right but an enabling one. Because it attenuates speech and noise in the direct path equally, it can raise the SNR at the eardrum only when the amplified path carries a better SNR than the direct path it displaces.

The spectral data show this pattern clearly. AI-SE produced nearly the same improvement in the amplified path of every device, about \SI{4}{dB} in the mid and high frequencies, but that improvement reached the eardrum only where the device output dominated the direct path. The benefit extended down to \SI{1.5}{kHz} with the tulip dome, to \SI{600}{Hz} with the closed coupling of the ANC-capable RIC, and to \SI{130}{Hz} once ANC was enabled, tracking the measured REOIG at each step. Because frequencies below \SI{1.5}{kHz} carry nearly half of the band importance for speech, extending the benefit downward increased its broadband value from 1.4 to \SI{3.5}{dB}. This is the behavior predicted by our earlier model~\citep{sabin2024} and by theoretical treatments of integrated ANC and noise reduction~\citep{serizel2010,serizel2013}, now observed in commercial hardware.

The practical implication is that the value of ANC in a hearing aid scales with the strength of its SNR-enhancing processing. Conventional processing produced modest added SNR here, leaving little for ANC to unlock; the additional improvement from AI-SE made the direct path the limiting factor. This is consistent with our model, which predicted substantial ANC benefit only when the device improves SNR by more than about \SI{4}{dB}~\citep{sabin2024}. As enhancement algorithms improve, the direct path will account for a growing share of the residual noise at the eardrum, and the benefit of attenuating it should grow accordingly. Conversely, evaluations of ANC in hearing aids without strong SNR-enhancing processing will understate its potential.

While these results validate a mechanism by which ANC can benefit speech-in-noise listening, the approach has several limitations. First, SII-weighted SNR at the eardrum is not equivalent to speech intelligibility. The mask-based SNR estimate does not characterize the artifacts or distortions that deep-learning enhancement can introduce~\citep{wang2017}, and whether the added SNR yields a corresponding improvement in speech recognition for listeners with hearing loss is an empirical question that only behavioral testing can answer. Second, this work used a single prescriptive target for a moderate sloping loss (N3). Our prior modeling~\citep{sabin2024} showed that ANC benefit decreases as prescribed gain increases, so we would expect smaller benefits for more severe losses and larger benefits for milder ones. Third, measurements were made on a single manikin; direct-path attenuation in real ears will vary with ear-canal geometry and the fit of the ear coupling.

% ---------------------------------------------------------------------------
\section{Conclusions}
% ---------------------------------------------------------------------------
In realistic loud-scene recordings on a manikin wearing commercial hearing aids, the SNR improvement produced by an AI speech-enhancement algorithm reached the eardrum only in the frequency range where the hearing aid's output dominated the direct path. Closed coupling widened that range relative to a vented fitting, and ANC widened it further, to the lowest frequencies measured. ANC by itself added almost no SNR (\SI{0.2}{dB}), but in combination with AI-SE it added \SI{1.8}{dB}. The full ANC-plus-AI-SE condition improved SII-weighted SNR by \SI{5.1}{dB} relative to the open ear, compared with \SI{1.6}{dB} for the \stdric{} with the same algorithm. These results provide physical-device evidence for the mechanism predicted by earlier modeling and motivate behavioral studies of integrated ANC and speech enhancement in hearing aids.

\section*{Competing interests}
A.T.S.\ is an employee of Bose Corporation and a scientific advisor to HearAdvisor. L.L.\ is an employee of Orka. Bose and Orka collaborated on manufacturing the hearing aids evaluated in this study.

% ---------------------------------------------------------------------------

\end{document}